\documentstyle[12pt]{article}

\begin{document}
\bibliographystyle{unsrt}

\begin{flushright} 
UMD-PP-96-94

April, 1996


\end{flushright}

\vspace{6mm}

\begin{center}

{\Huge \bf Neutron-Anti-Neutron Oscillation as a Test of Grand 
Unification\footnote{Invited talk presented at the International
workshop on proton decay and neutron oscillation at Oak Ridge National
Laboratory from March 28-30, 1996.}}

\vspace{6mm}

{\Large \bf  R.N. Mohapatra\footnote{ Work supported by the
 National Science Foundation Grant \#PHY-9119745 and a Distinguished
Faculty Research Award by the University of Maryland for the year 1995-96.}}\\

{\it{ Department of Physics }}\\
{\it{University of Maryland}}\\

{\it{ College Park, MD 20742 }}

\end{center}

\vspace{4mm}

\begin{center}
{\Large \bf Abstract}
\end{center}
\vspace{1mm}

We discuss the predictions for the neutron-anti-neutron ($N-\overline{N}$)
process in various supersymmetric and non-supersymmetric grand
unified theories. In particular it is pointed out 
that in a class of superstring
inspired grand unified theories (of $E_6$ type) that 
satisfy the constraints of gauge
coupling unification, breakdown of the $B-L$ symmetry occurs at an 
intermediate scale leading in turn to $\Delta B=1$ type R-parity violating
interactions naturally suppressed to the level of $10^{-5}$ to $10^{-7}$.
This in turn implies an $N-\overline{N}$ transition time of
order $10^{10}$ to $10^{11}$ sec. which may be observable in the next
generation of proposed experiments. These models also satisfy the
conditions needed for generating the cosmological baryon asymmetry
of the right order of magnitude for a restricted range of the parameter
space.

\newpage

\noindent{\Large \bf I. Introduction:}
\vspace{6mm}

The observed matter anti-matter asymmetry in nature is convincing enough as an
evidence for the existence of baryon number violation in the fundamental
interactions that describe physical processes. This is because of the
three conditions for generating this asymmetry laid down
originally by Sakharov in 1967: (i) existence of CP-violating  and
(ii) baryon number violating interctions plus (iii) the presence 
of out of thermal equilibrium conditions in the early universe. There are
also strong theoretical hints in favor of $\Delta B\neq 0$ 
interactions: for instance,
the standard model violates both baryon (B) and lepton (L) number 
via the triangle anomalies involving the electro-weak gauge bosons 
(although it conserves the linear combination $B-L$). The present
consensus however is that these baryon violating effects are too
weak to be observable in laboratory experiments.  Similarly, most 
extensions of the standard model also imply such interactions in the 
sector beyond the standard model. Perhaps more compelling is the argument
that the anomaly free gauge quantum number in most extensions of the
standard model is indeed the $B-L$ symmetry alluded to above. 
If the present indications for non-zero
neutrino mass from various terrestrial and extra terrestrial sources
hold up with time, the only sensible theoretical way to understand it is
to assume that the neutrino is a Majorana particle implying
that the lepton number is broken by vacuum by two units ($\Delta L=2$).
Since $B$ and $L$ appear in combination with each other, if lepton number
breaks by two units, there is no reason for baryon number not to break.
In fact this reasoning was first noted by Marshak and this 
author\cite{marshak} as a theoretical motivation for neutron-
anti-neutron oscillation. 

Once one accepts the existence of baryon number violating interactions,
it becomes of crucial importance to learn about the 
possible selection rules obeyed by them. As is quite
well-known\cite{weinberg}, the different selection rules probe new 
physics at different mass scales and therefore contain invaluable
information regarding the nature of short distance physics that is
otherwise inaccessible. Two of the most interesting
 selection rules are: one in which $B-L$ is conserved such as
the decay $p\rightarrow e^+ \pi^0$ (or $p\rightarrow \bar{\nu}_\mu K^+$
as in supersymmetric theories) and a second one which obeys $\Delta (B-L)
= 2$, such as $N-\bar{N}$ oscillation (which is the main theme of this
workshop). These two processes probe two
very different mass scales. To see this note that the process $p\rightarrow
e^+\pi^0$ arises from the operator $uude^-$ (or $QQQL$ in the $SU(2)_L\times
U(1)_Y$ invariant form) and it therefore scales like $M^{-2}$ where M
denotes the mass scale where the interaction originates. The present limits
on proton lifetime then imply that $M\geq 10^{15}$ GeV or so. On the
other hand, $N-\bar{N}$ oscillation arises from the operator of the
form $u^cd^cd^cu^cd^cd^c$ which scales like $M^{-5}$. The limits
on nonleptonic $\Delta B=2$ nuclear decays or the $N-\bar{N}$ oscillation
time from the ILL experiment\cite{ILL}, implies that $M\geq 10^5$ GeV or so.
Thus $N-\bar{N}$ oscillation has the additional properties that it also
provides complementary probes of new physics near the TeV scale.

There is however as yet no laboratory evidence for any kind of $\Delta B\neq 0$
process. The first generation of experiments searching for evidence of
baryon number violation have all reported their results as lower
limits on the partial life times for the various decay modes of the proton
(at the level of roughly $10^{32}$ to $10^{33}$ years). Those results
have already had the important implication that the minimal non-supersymmetric
$SU(5)$ model is ruled out as a grand unification theory. There are currently
two experimental efforts to improve the discovery potential for proton decay
to the level of $10^{34}$ years. These are the Super-Kamiokande\cite{stone} 
and ICARUS\cite{cline}
experiments. To go beyond that would require a major innovation in
experimental methods.

There is however encouraging news from the $N-\bar{N}$ oscillation front
in this regard. As is well-known\cite{rnm}, the existence of neutron-
anti-neutron oscillation inside nuclei leads to baryon instabilty
which can also be probed in the proton decay searches (e.g. the disppearance
of oxygen nuclei in water detectors). One can then use simple scaling
arguments to relate the nuclear instability life time($\tau_{nucl}$)
to the $N-\bar{N}$ oscillation time ($\tau_{N-\bar{N}}$) For more
reliable nuclear physics calculations, see \cite{alberi}:
\begin{eqnarray}
\tau_{nucl}\simeq \left({{\tau_{N-\bar{N}}}\over{6.6\times 10^{6}~sec}}
\right)^2\times 10^{30}~yrs.
\end{eqnarray}
From this equation we see that a measurement of $\tau_{N-\bar{N}}$ to the
level of $10^{10}$ sec. (as is contemplated by the Oak Ridge group\cite{yuri})
 would correspond to probing baryon 
instability to the level of almost $10^{37}$
yrs. This will take us far into the uncharted domain of baryon 
non-conservation not easily accessible in other experiments(albeit
in a very special non-leptonic channel). This may be one of the strongest
arguments for undertaking such an experiment. In this article, I will
discuss elegant and plausible theoretical models that provide additional
arguments in favor of conducting such an experiment since this can be a 
very useful way to discriminate between various grand unification theories.

This paper is organized as follows: in sec.II, the general theoretical
arguments for $N-\bar{N}$ oscillation based on gauged $B-L$ symmetry
are outlined; the predictions for $\tau_{N-\bar{N}}$ in non-supersymmetric
theories and supersymmetric theories are given in sec.III and IV respectively;
in sec.V, it is shown how baryon asymmetry can be generated in an $[SU(3)]^3$
string inspired SUSY GUT model which predicts observable $\tau_{N-\bar{N}}$;
in sec.VI, some concluding remarks are presented.

\vspace{6mm}
\noindent{\Large \bf II. Local $B-L$ symmetry and $N-\bar{N}$ oscillation:}
\vspace{6mm}

As already mentioned, in the standard model, $B-L$ is an anomaly free global
symmetry. However, it is not a gaugeable symmetry since it is not cubic
anomaly free. This fact is connected with whether neutrino mass is zero
or not. In the standard model neutrino mass vanishes because the
right handed neutrino is not included in the spectrum; it is also the 
absence of $\nu_R$ that prevents $B-L$ symmetry from being a gaugeable
symmetry. In order to generate neutrino mass, we must add $\nu_R$
to the spectrum of fermions in the standard model. As soon as this is
done, $B-L$ becomes cubic anomaly free and becomes a gaugeable symmetry.
This also incidentally restores quark-lepton symmetry to particle physics.
The natural gauge symmetry of particle physics then becomes the left-right
symmetric gauge group $SU(2)_L\times SU(2)_R\times U(1)_{B-L}$\cite{lrs}
which then not only explains the smallness of neutrino mass but it also
makes weak interactions asymptotically parity conserving. It was noted
in 1980\cite{marshak,davidson} that in formula for electric charge in
the left-right symmetric model is given by:
\begin{eqnarray}
Q=I_{3L}+I_{3R}+{{B-L}\over{2}}
\end{eqnarray}

It follows from this equation\cite{mm} that since $\Delta Q = 0$,
at distance scales where $\Delta I_{3L}=0$, we have the relation
\begin{eqnarray}
\Delta I_{3R}=-{{1}\over{2}}(B-L)
\end{eqnarray}

Clearly, the violation of lepton number which leads to a Majorana mass
for the neutrino is connected with the violation of right-handed iso-spin
$I_{3R}$. The same equation also implies that for processes where no
leptons are involved, it can lead to purely baryonic processes where 
baryon number is violated. In nonsupersymmetric theories, the simplest such
process is neutron-anti-neutron oscillation since $u^cd^cd^cu^cd^cd^c$
is the lowest dimensional baryon number violating operator that conserves
color, electric charge and angular momentum 
and does not involve any lepton fields. 
(The situation is different in supersymmetric 
theories as we will see in the next section.) This equation implies
a deep connection between the Majorana mass for the neutrino and the existence
of neutron-anti-neutron oscillation. Of course whether  $N-\bar{N}$ 
transition appears with an observable strength  depends on the details
of the theory such as the mass spectrum, value of mass scales etc.

Before proceeding further, a few words about the notation: Let us call
$G_{N-\bar{N}}$ as the strength of the six quark amplitude; 
$\delta m_{N-\bar{N}}$
as the transition mass for neutron-anti-neutron transition and 
$\tau_{N-{\bar{N}}}=h/2\pi \delta m_{N-\bar{N}}$ 
where $h$ is Planck's constant.
We hasten to clarify that while theories with local $B-L$ symmetry provide
a natural setting for the neutron-anti-neutron oscillation to arise, it is
possible to construct alternative models where one can have $N-\bar{N}$
oscillation. In such models however, the strength for this process is
completely unrelated to other physics making them quite adhoc.

\vspace{6mm}
\noindent{\Large \bf III. Predictions for $\tau_{N-\bar{N}}$ in
non-supersymmetric unified theories:}
\vspace{6mm}

There were many models for neutron-anti-neutron oscillation discussed in
the early eighties\cite{rnm}; most of these models are in the context of
nonsupersymmetric higher unified theories. Here I present the simplest of them
and summarize the general status of $\tau_{N-\bar{N}}$ 
transition in all these models
in Table 1. 

We will consider the gauge group $SU(2)_L\times SU(2)_R\times
SU(4)_c$ which was suggested by Pati and Salam\cite{lrs} in 1973.
The recognition that $SU(4)_c$ contains the $B-L$ symmetry
and has the potential to explain neutrino mass and applications
to N-$\bar{N}$ oscillation came in the papers of Marshak and this author
\cite{marshak,mm}. In order 
to obtain neutron-anti-neutron oscillation process,
the gauge symmetry breaking of the model has to be broken
by the Higgs multiplets as in Ref.\cite{mm} 
i.e. a bidoublet $\phi(2,2,1)$,
and a pair of triplets $\Delta_L(3, 1 , \bar{10})+\Delta_R(1, 3, \bar{10})$.
This set of Higgs multiplets was different from the one originally
used in Ref.\cite{lrs} and brought out the physics of the model
in a very clear manner.
The quarks and leptons are assigned to representations as follows:
$Q_L(2,1,4)+Q_R(1,2,4)$. Here leptons are considered as the fourth color.
The allowed Yukawa couplings in the model are given by:
\begin{eqnarray}
L_Y=y_q \bar{Q}_L\phi Q_R + f(Q_LQ_L\Delta_L+Q_RQ_R\Delta_R) + h.c.
\end{eqnarray}
Here we have omitted all generation indices and also denoted the couplings
symbolically omitting charge conjugation matrices, Pauli matrices etc.
The Higgs potential of the model can be easily written down; the term
in it which is interesting for our purpose is $\lambda 
\epsilon^{ijkl}\epsilon^{i'j'k'l'}\Delta_{L,ii'}\Delta_{L,jj'}\Delta_{L,kk'}
\Delta_{L,ll'} + L\rightarrow R + h.c.$.          

In order to proceed towards our goal of estimating the strength of $N-\bar{N}$
oscillation in this model, we first note that the original gauge symmetry
here is broken by the vev $\langle \Delta_{R,44} \rangle=v_{B-L}
 \neq 0$ to the standard model gauge group. The diagram of Fig.1 then
leads to the six quark effective interaction below the scale $v_{B-L}$ of
the form $u_Rd_Rd_Ru_Rd_Rd_R$ with strength $\lambda f^3 v_{B-L}/ 
M^6_{\Delta_R}$. For the scale $v_{B-L}$ and $M_{\Delta_R}$ of order 100
TeV and for $h\approx \lambda \approx 10^{-1}$ this
will lead to a strength for the six quark amplitude of about $10^{-29}$ 
GeV$^{-5}$. In order to convert it to $\delta m_{N-\bar{N}}$, we need 
the three quark "wave function" of the neutron at the origin. This has been
estimated by various people\cite{pasu} and usually yields a factor of about
$10^{-4}$ or so. Using this , we expect $\tau_{N-\bar{N}}\simeq 6\times 
10^{8}$ sec. This is however only an order of magnitude estimate since the 
true value of the parameters that go into this estimate is unknown. But the
main point that this example makes is 
that there exist very reasonable theories 
where neutron-anti-neutron oscillation is observable. Note that this model is
a completely realistic extension of the standard model with many intersting
features such as the smallness of neutrino mass naturally explained etc. 

A natural question to ask at this point is whether there are grand
unified theories where observable $N-\bar{N}$ oscillation can be expected.
In simple nonsupersymmetric extensions of $SU(5)$ model, it is easy to
show that\cite{book} $N-\bar{N}$ transition amplitude is 
very highly suppressed. Let us therefore consider the $SO(10)$ model which
contains the gauge group $SU(2)_L\times SU(2)_R\times SU(4)_c$ . All the
ingredients for a sizable $N-\bar{N}$ to exist are present in the model
except that the scale  $v_{B-L}$ is constrained by gauge coupling unification.
This question was studied in detail in Ref.\cite{chang} and a scenario of 
symmetry breaking was isolated where one could get a value for $v_{B-L}\simeq
100$ TeV. This would therefore lead to an observable $\tau_{N-\bar{N}}$
oscillation as before. The only problem is that in a low $SU(4)_c$ scale
model, one has to introduce iso-singlet fermions to lift the degeneracy
between quark and charged fermion masses implied by $SU(4)_c$ symmetry.
While this procedure is quite harmless in partial unification models,
it effects gauge coupling unification in a model such as $SO(10)$. This 
question has not been discussed yet in such models. For situation in
other non-SUSY GUT theories, see Table 1.

\vspace{6mm}
\noindent{\Large \bf IV. R-parity violation and $N-\bar{N}$ oscillation:}
\vspace{6mm}

The particle physics of the nineties has perhaps a different "flavor"
(set of prejudices ?) than the eighties.
It is now widely believed that supersymmetry is an essential ingredient
of physics beyond the standard model with supersymmetry breaking scale
around a TeV in order to explain the origin of 
electroweak symmetry breaking. Furthermore if one believes that
supersymmetry is the low energy manifestation of the superstring theories,
then to the usual renormalizable Lagrangian of the supersymmetric theory,
one must add non-renormalizable terms which are the low energy remnants
of superstring physics. In the discussion of this section, we will
use both these ingredients. 

A simple way to explain supersymmetric theories is to note that
corresonding to every particle there is a super partner (spin half
partner for a gauge boson or Higgs boson 
and spin zero partner for a fermion with identical
internal quantum numbers in both cases) 
and there are a large number of relations 
between the coupling constants of the theory. In this article, we
will denote the super partners of quarks and leptons by $\tilde{q}$
and $\tilde{L}$ respectively; 
super partners of $W$ and $Z$ bosons by $\tilde{W}$
and $\tilde{Z}$ etc. The extension of the standard model to include
supersymmetry is under extensive investigation right now both 
from theoretical and experimental side.

One troubling aspect of the minimal supersymmetric 
extension of the standard model (MSSM) is that
it allows for lepton and baryon number violating interactions with
arbitrary strengths. This in a sense is a step backward from the 
standard model which automatically ensured that both baryon and
lepton numbers are conserved to an extremely high degree
as is observed. A simple way to
see the origin of such terms is to note that $\tilde{L}$ which is the
superpartner of the lepton doublet is exactly like a Higgs boson of
the standard model except that it carries lepton number. 
But we know that the Higgs doublet of the standard model couples to
quarks; similarly the $\tilde{L}$ field also
couples to quarks as in the standard model: $Q\tilde{L}d^c$; but this
clearly violates lepton number by an arbitrary amount. Similar terms
can be written down which violate baryon number also with arbitrary strength.
These are the so-called R-parity violating
interactions. There exist very stringent upper limits
on the various R-parity violating coupligs\cite{rv} which range
anywhere from $10^{-4}$ to $10^{-8}$ depending on the type of
selection rules they break. 
Since the main reason for believing in supersymmetry
is that it improves the naturalness of the standard model, it will be 
awkward to assume that the MSSM carries along with it the above fine-tuned
couplings without any fundamental assumptions. 

The general attitude to this problem is
that when the MSSM is extrapolated to higher scales, new symmetries
will emerge which either forbid the R-parity violating couplings or
suppress it in a natural manner. A concrete proposal in this direction
proposed some time ago\cite{rm1} is 
that at higher energies the the gauge symmetry
becomes bigger and includes $B-L$ as a subgroup. It is well-known
that the $B-L$ symmetry
 is also important in understanding the smallness of the neutrino mass;
therefore is not a completely new symmetry custom-designed
only to solve the R-parity problem. 
It is easy to see that in the symmetric phase of a theory
containing $B-L$ local symmetry, R-parity
is conserved since $R=(-1)^{3(B-L)+2S}$. This however is not the
end of the story since the $B-L$ must be a broken symmetry at low
energies and if the $B-L$ symmetry is broken by the vev of a scalar field
which carries odd $B-L$, then R-parity is again broken at low 
energies\cite{rm1}. Examples of  theories where R-parity is
broken by such fields abound- the string inspired $SO(10)$ and $E_6$
being only two of them. On the other hand there are also many theories
where $B-L$ is broken by fields with even $B-L$ values.
In these models\cite{rm2}, 
R-parity remains an exact symmetry, as is required if
supersymmetry has to provide a cold dark matter particle. It remains to
be seen whether these latter class of models can arise from some higher
level compactification of superstring theories.

In this paper we focus on the first class of theories since it has been
shown that they can arise from string theories in different compactification
schemes. In this class of theories, R-parity breaking interactions arise
once $B-L$ symmetry is broken. It is then easy to see that to suppress the
R-parity breaking interactions to the desired level, $B-L$ breaking must
occur at an intermediate scale\cite{lee} 
than at the GUT scale as is quite often done.
The reason why all this is of interest to us is
that while pure lepton number violating 
processes in these classes of models are likely to be highly suppressed, the
$\Delta B=2$ processes such as neutron-anti-neutron oscillation may arise at
an observable rate. To see what kind of restrictions on R-parity breaking
couplings are implied by the present lower limits 
on $N-\bar{N}$ transition time, let us start by writing down
the general structure of R-parity violating
interactions in the MSSM:
\begin{eqnarray}
W_{RP}=\lambda_{ijk} L_i L_j e^c_k+\lambda'_{ijk} Q_i L_j d^c_k +
\lambda''_{ijk} u^c_i d^c_j d^c_k
\end{eqnarray}
The coupling relevant in the discussion of neutron-anti-neutron oscillation
is the $\lambda''$\cite{zw}. 
Due to the color structure of the coupling, it cannot
lead to $N-\bar{N}$ oscillation in the tree level and one has to invoke
electroweak loop effects. This has been studied in detail in the recent
paper of Goity and Sher\cite{goity}. They conclude that the dominant
contribution arises from the $u^cd^cb^c$ type coupling in conjunction
with a box diagram that changes $dd\rightarrow bb$ and has the strength
(see Fig.2):
\begin{eqnarray}
G_{N-\bar{N}}= {{6\alpha^2_{wk}m_{\tilde{W}}m^2_b V_{ub}\tilde{V}_{ub}
\lambda''^2_{123}}\over{M^8_{\tilde{b_L}}}} GeV^{-5}
\end{eqnarray}
The $V_{ub}$ and $\tilde{V}_{ub}$ above refer to the $ub$ mixing angles
in the quark and squark sector. The rest of the notation is self 
explanatory. The value of $\tilde{V}_{ub}$ is not known.
In order to estimate the transition time for neutron-anti-neutron oscillation,
we have to multiply by the wave function effect i.e. $|\psi(0)|^2$:
\begin{eqnarray}
\delta m_{N-\overline{N}}=G_{N-{\overline{N}}} |\psi(0)|^2~ GeV
\end{eqnarray}
Using the value for $|\psi(0)|^2\simeq 3\times 10^{-4}$ $GeV^6$
from Ref.\cite{pasu},
we get
\begin{eqnarray}
\delta m_{N-\overline{N}}\simeq 5\times 10^{-22}\lambda''^2_{123}
 \left({{100 GeV}\over{M_{sq}}}\right)^6 GeV
\end{eqnarray}
The ILL lower bound on $\tau_{N-\bar{N}}\geq .8\times 10^8$ sec. can be
translated into an upper bound on $\lambda''_{123}\leq 4\times 10^{-6}$. 
There are uncertainties in this estimate
coming from the value of squark mixings as well as the values
of squark masses. Our goal will be to seek grand unified theories where
values of $\lambda''$ in the general ball-park $10^{-6}$ to $10^{-7}$
are predicted so that one may
confidently argue that those models provide a good motivation for
carrying out the neutron oscillation experiment.

We will be guided in our choice of the models by the heterotic superstring
theory compactified either fermionically or via the Calabi-Yau manifolds.
It turns out that complete breakdown of the gauge symmetry in these cases
automatically imply that R-parity, which is an exact symmetry  above the
GUT scale breaks down. Our goal will be to study the prediction of the
strength R-parity violating interactions in these models consistent with
the idea of gauge coupling unification. We will discuss two classes of
theories: one based on the gauge group $SO(10)$ and another on $[SU(3)]^3$.
In both cases we will restrict ourselves to only those Higgs representations
allowed by the superstring compactification guidelines.

\vspace{6mm}
\noindent{\Large \bf V. Spontaneous breaking of R-parity in string inspired
$SO(10)$ model:}
\vspace{6mm}

 In the $SO(10)$ model,
 the matter fields belong to the spinor {\bf 16}-dimensional
representations whereas the Higgs fields will belong to {\bf 45}, {\bf 54},
{\bf 16}+$\overline{\bf 16}$ {\bf 10}-dim 
representations as is suggested by recent studies of level
two models\cite{lykken}. The symmetry breaking in these models is achieved as
follows: The vev of the {\bf 45} and {\bf 54}-dim fields 
break the $SO(10)$ symmetry down to $SU(3)_c\times SU(2)_L\times SU(2)_R\times
U(1)_{B-L}$ which is broken down to the standard model by the $\tilde{\nu^c}$
component of {\bf 16}+$\overline{\bf 16}$ acquiring vevs. The question we now
ask is what is the strength of $\Delta B=1$ 
R-parity violating terms at low energies. Since the $\tilde{\nu}^c$
field has $B-L=1$, it will induce the $\Delta B=1$ terms at low energies.
First point to note that they do not arise from renormalizable terms
in the Lagrangian but rather only from the mass suppressed nonrenormalizable
terms in the $SO(10)$ model. This imlies that they are automatically
suppressed. The relevant terms are of the form ${\bf 16}_H {\bf 16}_m 
{\bf 16}_m {\bf 16}_m/ M_{Pl}$. When $\tilde{\nu^c}$ vev is turned on,
these type of terms lead to terms of type $QLD^c$, $LLE^c$ as well as
$U^cD^cD^c$. Their strength will be given by $\langle \tilde{\nu^c}\rangle/
M_{Pl}$ and will therefore depend on the scale of $B-L$ breaking, which
in turn is tied with the gauge coupling unification. Important point to
note is that all the above terms have the same strength as a result
of which a combination of the 
$QLD^c$ and the $U^cD^cD^c$ terms at the tree level will lead to proton
decay with strength $\simeq {{\alpha_W m_{\tilde{w}}}\over{4\pi M^2_{sq}}}
\left({{\langle \nu^c\rangle}\over{M_{Pl}}}\right)^2$. The present limits
then imply that  $\langle \nu^c\rangle / M_{Pl} \leq 10^{-12}$.
This automatically implies that the effective $\lambda''$ type terms are
also of this order leading to unobservable amplitudes for $N-\bar{N}$
transition.

\vspace{6mm}
\noindent{\Large \bf VI. Observable $N-\bar{N}$ oscillation
in $[SU(3)]^3$ model:}
\vspace{6mm}

Let us now turn to the superstring inspired $[SU(3)]^3$ type models.
The matter multiplets in this case belong to representations
$\psi\equiv({\bf 3, 1, 3})$, $\psi^c\equiv({\bf 1,\overline{3}, \overline{3}})$
 and $F\equiv({\bf \overline{3}, 3, 1})$ representations. 
The particle content of these representations can be given by:
$\psi=\left( u, d,g \right)$, $\psi^c=\left(u^c,d^c,g^c \right)$,
\begin{eqnarray}
F=\pmatrix{
H^0_u & H^+_d & e^+\cr
H^-_u & H^0_d & \nu^c \cr
e^-   & \nu   & n^0 \cr}
\end{eqnarray}
$\psi$ and $\psi^c$ denote the quark multiplets and $F$ denotes the
leptonic multiplets. The Higgs fields will belong to $F$-type representations
and will be denoted by $H$ and $\bar{H}$ respectively.
 The gauge invariant couplings are then given as in the following
 superpotential:
\begin{eqnarray}
f\psi \psi^c H+ f'(\psi \psi \psi+\psi^c\psi^c\psi^c)+f''\psi\psi^c F
+h_1 F^3+h_2 H^3 + h_3\bar{H}^3+ ...
\end{eqnarray}
where we have suppressed the generation indices.
These terms are of course enormalizable.
Again as in the case of the SO(10) model, the R-parity violating terms
arise once the $\tilde{\nu^c_H}$ vev is inserted in the above operators. Again,
as before, $\Delta B\neq 0$ terms will be induced by
purely renormalizable terms
thru tree diagrams of type shown in Fig.3. They lead to $u^cd^cd^c$
type terms\cite{valle} . It is these type of terms that are dominant and
their strength can be estimated to be $f f'\langle \tilde{\nu^c}\rangle/ 
{\langle n^0\rangle}$.  The strength of $\Delta B\neq 0$
R-parity violating terms are dictated by gauge coupling unification.

Let us now see the constraints of proton decay on the couplings in this
 model. To see this, let us recall the superpotential in the above 
equation. Note that proton decay involves the couplings $f'f''$ whereas
$\Delta B=1$ non-leptonic terms involve $ff'$. Therefore unlike the
SO(10) case, the two processes are decoupled from each other and we
can suppress proton decay by imposing a symmetry that forbids the
$f''$ term but not the $f$ or $f'$ terms.

Let us now proceed to discuss the constraint of gauge coupling unification
on the $B-L$ breaking scale in these models. It turns out that if 
 we assume that $[SU(3)]^3$ breaks down to $SU(3)_c\times SU(2)_L
\times SU(2)_R\times U(1)_{B-L}$ at the GUT scale by the vev of the $n^0$
field, the spectrum of particles below it is same as 
for the $SO(10)$ case. We keep one additional color octet multiplet
below $M_U$. The one and two loop unification in this case have been
studied recently\cite{bastero} and the result is that one finds $M_U\simeq
10^{18}$ GeV and $M_{B-L}\simeq 10^{13}$ GeV or so. The one loop unification
graph is shown in Fig.4.
 We see from the discussion in the above sections that operators 
of type $u^cd^cd^c$ are induced with strength of order $\lambda f$
where $\lambda \simeq 10^{-5}$ as determined by the unification analysis
and $f$ is an unknown parameter (which could be assumed to be
of order $10^{-1}$). This leads to $\lambda''\simeq 10^{-7}$ or so.
It can lead to observable neutron-anti-neutron
oscillation with $\tau_{N-\bar{N}}$ of order $10^{10}$ sec. We
hasten to note that due to the unknown coupling $f$ in the six-quark superfield
operator, we cannot make an exact prediction; but given the uncertainties
in the parameters, the neutron-anti-neutron oscillation time could
be somewhere between $10^8$ to $10^{10}$ sec. This is clearly accssible
to the proposed Oak Ridge experiment which plans
 to search for neutron-anti-neutron
oscillation upto a sensitivity of $10^{10}$ to $10^{11}$ sec.\cite{yuri}. 
This should
therefore throw light on the nature of this class of grand unified theories.

\vspace{6mm}
\noindent{\Large \bf VII. Baryogenesis in the $[SU(3)]^3$ model:}
\vspace{6mm}

In the section we present a brief outline of a scenario for baryogenesis
in the $[SU(3)]^3$ model discussed above. The reason for this is that
the nature of the selection rule for baryon number non-conservation
and the possibility of baryogenesis in the early universe are intimately 
linked. Very crudely this connection can be stated as follows: the higher
the dimensionality of the $\Delta B\neq 0$ operator, the lower is the 
temperature of its thermodynamic decoupling from the rest of the universe.
Since before the decoupling temperatures is reached 
such processes can always erase any preexisting baryon asymmetry, there 
is a close connection between the mechanism for baryogenesis and the
nature of baryon non-conservation. Clearly, since the N-$\bar{N}$ 
transition operator has dimension nine, it remains in equilibrium to
very low temperatures and one must be careful.

We contemplate the following scenario for baryogenesis, where the lepton
asymmetry of the universe is generated at temperatures of order $10^9$
GeV or so below the temperature for inflation reheating. This lepton symmetry
is converted to the baryon asymmetry due to sphaleron effects \cite{kuzmin}
as suggested in Ref.\cite{fuku}. We now have to make sure that the 
$\Delta B=1$ interaction is out of thermal equilibrium during the time
when the sphalerons active in transforming the lepton number into baryon
number i.e. from $10^{9}$ GeV down to $100$ GeV. 
In Fig.5, we have plotted the rates for the $\Delta B=1$ process
and the Hubble expansion rates for various values of the $\lambda''$
coupling. It appears that only for $\lambda''\leq 10^{-7}$ or so,
the conditions are favorable for baryogenesis. One could also treat
this as a crude upper bound on the magnitude of the $\Delta B=1$
interaction from the baryogenesis consideration. This corresponds
to a $\tau_{N-\bar{N}}\simeq 10^{10}$ sec. It is interesting that
this is the range being expected from $[SU(3)]^3$ type theories and
is also measurable in the $N-\bar{N}$ experiment being planned.

\vspace{6mm}
\noindent{\Large \bf VIII. Conclusion:}
\vspace{6mm}

In conclusion, it is clear that a dedicated search for neutron-anti-neutron
oscillation to the level of $10^{10}$ sec sensitivity is going to prove
extremely valuable in our understanding of physics beyond the standard model.
A non-zero signal would rule out many grand unified theories such as the simple
non-supersymmetric $SU(5)$ and $E_6$, 
supersymmetric $SO(10)$ models etc. and will be a 
strong indication in favor of a string inspired supersymmetric
$E_6$ or $[SU(3)]^3$ type model. A negative
signal to this level would imply restrictions on the baryogenesis scenarios
and the accompanying particle physics models. A positive signal would
also yield valuable information on the violation of equivalance principle
between particle and anti-particle.

I would like to thank the Oak Ridge National laboratory for hospitality
during the workshop.

\begin{center}

Table 1

\begin{tabular}{|c||c||c|} \hline
GUT model  & Is $N-\bar{N}$ observable? & Implications \\ \hline 
(NON SUSY)   &      &     \\ \hline
$SU(5)$   &   No  & $\Delta(B-L)=0$ \\ \hline
$SU(2)_L\times SU(2)_R\times SU(4)_c$ & Yes & $M_c \simeq 10^5$ GeV\\ \hline
Minimal SO(10) & No &   \\ \hline
$E_6$   &  No  &    \\ \hline

(SUSY GUT) &    &   \\ \hline

$[SU(3)]^3$  &  Yes  & Induced breaking of R-parity \\ \hline
$SO(10)$    &   No   &    \\ \hline
\end{tabular}

\end{center}

\noindent{\bf Table Caption:} This table summarizes the  observability
of neutron-anti-neutron oscillation in various GUT models.

\vspace{6mm}

\noindent{\bf Figure Caption:}

\noindent{\bf Fig 1:} The Feynman diagram that leads to $N-\bar{N}$
oscillation in the $SU(2)_L\times SU(2)_R\times SU(4)_c$ model.

\vspace{3mm}

\noindent{\bf Fig 2:} The diagram responsible for $N-\bar{N}$ oscillation
in models with R-parity breaking.

\vspace{3mm}

\noindent{\bf Fig 3:} The origin of the $u^cd^cd^c$ vertex in $[SU(3)]^3$
type model at low energies.

\vspace{3mm}

\noindent{\bf Fig 4:} The running of gauge couplings in the one loop 
approximation in models with intermediate scales and unification at the
string scale.

\vspace{3mm}

\noindent{\bf Fig 5:} The comparision of the rates for baryon number
violating processes in R-parity broken models with the Hubble expansion rate.

\end{document}